# Kirigami-based Flexible Metasurface with Reconfigurable Intrinsic Chirality from Zero to Near-unity

Yiyi Yao[1], Shijie Kang[1,2], Aoning Luo[1], Jiusi Yu[1], Ken Qin[1], Xiexuan Zhang[1], Jiayu Fan[1], Xusheng Xia[3,4,*], Haitao Li[1,*], Xiaoxiao Wu[1,2,*]

[1]Modern Matter Laboratory and Advanced Materials Thrust, The Hong Kong University of Science and Technology (Guangzhou), Nansha, Guangzhou 511400, China

[2]Low Altitude Systems and Economy Research Institute, The Hong Kong University of Science and Technology (Guangzhou), Nansha, Guangzhou 511400, Guangdong, China

[3]State Key Laboratory of Advanced Glass Materials, Wuhan University of Technology, Wuhan, 430070, China

[4]School of Material Science and Engineering, Wuhan University of Technology, Wuhan 430070, China

*To whom correspondence should be addressed. E-mails: xsxia@whut.edu.cn (X. Xia), haitaoli@hkust-gz.edu.cn (H. Li), xiaoxiaowu@hkust-gz.edu.cn (X. Wu).

## Abstract

Chiral responses in electromagnetic metasurfaces are typically categorized as extrinsic, resulting from asymmetric interactions between the structure and incident waves, and intrinsic, arising from three-dimensional symmetry breaking of the unit cell. However, most existing metasurface designs target only one type of chirality and lack a unified, continuously tunable platform for broader chiroptical control. To address this limitation, the designed kirigami-based flexible metasurface is proposed for dynamic, continuous modulation of chirality, which expands the control scope to both extrinsic and intrinsic chiral responses within a single, reconfigurable platform. Initially, the unfolded metasurface exhibits extrinsic chirality under oblique incidence. By introducing well-designed kirigami-based cuts and folds, the metasurface transitions from a planar and achiral configuration to a three-dimensional chiral geometry that breaks the mirror symmetry, thereby exhibiting tunable intrinsic chirality and asymmetric extrinsic chirality. As the folding angle increases, the resulting deformation enables continuous tuning of the chiral response, with circular dichroism and its asymmetry under oblique incidences progressively increasing and reaching pronounced levels across the X-band. Our work provides a lightweight, easy-fabricated, and mechanically reconfigurable metasurface, which offers strong potential for future development in adaptive photonic systems and advanced chiroptical technologies.

## 1. Introduction

Chirality, a fundamental property where objects cannot be superimposed onto their mirror images through rotation or translation[1-2], is ubiquitous across all scales, from galaxies to molecules like sugars and DNAs[3]. Chiral materials exhibit distinct responses to left- and right-handed circularly polarized waves, manifesting as circular dichroism (CD) and optical activity[4]. However, natural chiral materials often display weak and inflexible responses, limiting their practical utility, especially in long wavelength cases. Electromagnetic (EM) metasurfaces, engineered microstructures that manipulate waves at subwavelength scales, have revolutionized applications like wavefront shaping[5-7], lensing[8-9], holography[10-12], and vortex beam generation[13-15]. By integrating chirality into metasurface design, researchers have explored two primary types: extrinsic chirality, arising from asymmetric wave-structure interactions under specific illumination conditions[16-21], and intrinsic chirality, stemming from symmetry-breaking within the unit cell, enabling robust responses even under normal incidence[22-26].

While significant advances have been made in designing metasurfaces for either extrinsic or intrinsic chirality—through techniques like tuning incidence angles or introducing asymmetric geometries, these approaches typically address only one type of chirality and lack unified control. Moreover, achieving continuous tunability of chiral responses on a single platform remains elusive, hindering applications requiring versatile, multi-state chiral control.

In recent years, kirigami and origami, traditional arts of cutting and folding, have emerged as innovative approaches in materials science, offering reconfigurable frameworks for metasurface design[27-32]. Kirigami, in particular, enables the transformation of planar structures into complex three-dimensional (3D) forms through mechanical deformations like stretching, folding, or bending. Unlike conventional tuning methods reliant on electronic components[33-34] or active materials[35-36], kirigami-based metasurfaces provide a passive, mechanically driven approach to modulate geometry and electromagnetic properties. Benefits include low power consumption, ultrathin profiles, and ease of fabrication, making them ideal for dynamic photonic systems[37-39]. Despite these advances, most kirigami-inspired metasurfaces achieve only binary chiral transitions, switching between achiral and chiral states or between opposite chiralities[40-42]. Such limitations restrict their ability to meet demands for precise, multi-state chiral control.

In this work, we introduce a kirigami-based metasurface enabling continuous tunability of intrinsic chiral responses through adjustable folding angles. This design, based on kirigami, which bridges the extrinsic and intrinsic chirality, facilitates a seamless transition of the metasurface from an achiral state (CD ~ 0) to near-unity CD (CD ~ 0.90) for normally incident EM waves, as shown in Fig. 1(a). In its planar state, the metasurface exhibits strong extrinsic chirality under oblique incidence, while increasing folding angles through kirigami induces symmetry breaking, yielding robust and tunable intrinsic chirality. Experimental results, corroborated by simulations, reveal that asymmetric excitation of toroidal and electric dipole moments under circular polarizations drives this response, with reactive helicity density (RHD) confirming the link between geometric reconfiguration and chirality. After folding, the kirigami-based metasurface also exhibits extrinsic chirality that is strongly asymmetric with respect to opposite angles of incidence. In general, our kirigami-based metasurface can serve as an ultrathin, lightweight, and dynamically tunable platform of chiroptical responses that paves the way for advanced reconfigurable photonic devices.

## 2. Results and Discussion

The metasurface is composed of a polyimide (PI) film and periodically arranged copper split ring resonators (SRRs). Fig. 1(a) illustrates the mechanism of our metasurface. Initially, when the metasurface is unfolded, the two SRRs are coplanar and arranged with approximate mirror symmetry with respect to the *xy*-plane, although the PI substrate introduces a slight asymmetry. As a result, the system exhibits no intrinsic chirality, although extrinsic chirality may appear under oblique incidence (see Supporting Information Note S1). By cutting and folding the structure along the dashed lines shown in Fig. 1(b), the originally coplanar SRRs become staggered. This deformation significantly breaks mirror symmetry along the *z*-axis and induces a strong chiral response. Fig. 1(c) illustrates the details of the unit cell of the metasurface and its specific deformation process, which is achieved through cutting and folding. The folding angle $\theta$ is defined as the dihedral angle between the folded surface and the *xy*-plane (see Supporting Information Movie S1 for animation of the folding with different folding angles). As denoted in Fig. 1(b), the geometric parameters of the metasurface are defined as follows: the width $W = 25$mm, the length $L = 18$mm; and other geometry parameters: $a = 12$mm; $b = 4$mm; $c = 8$mm; $d = 6$mm; $g_1 = 3$mm; $g_2 = 1$mm and $t_1 = 0.5$mm; the polyimide (PI) thickness $t_0 = 0.1$ mm, and the copper thickness $t_2 = 0.035$ mm. With such kirigami-inspired cuts, when the metasurface is unfolded, it is still achiral and allows the equal transmissions of both left-handed circularly polarized (LCP) and right-handed circularly polarized (RCP) waves under normal incidences. As shown in Fig. 1(a), when the folding angle $\theta$ gradually increases, it begins to exhibit an enhanced chiral response and ultimately enables highly efficient transmission of LCP incidences while effectively suppressing RCP transmissions at larger folding angles. This evolution demonstrates that the chiral response under normal incidence is highly dependent on the folding angle, enabling a continuous transition of the metasurface from an achiral to a highly chiral state.

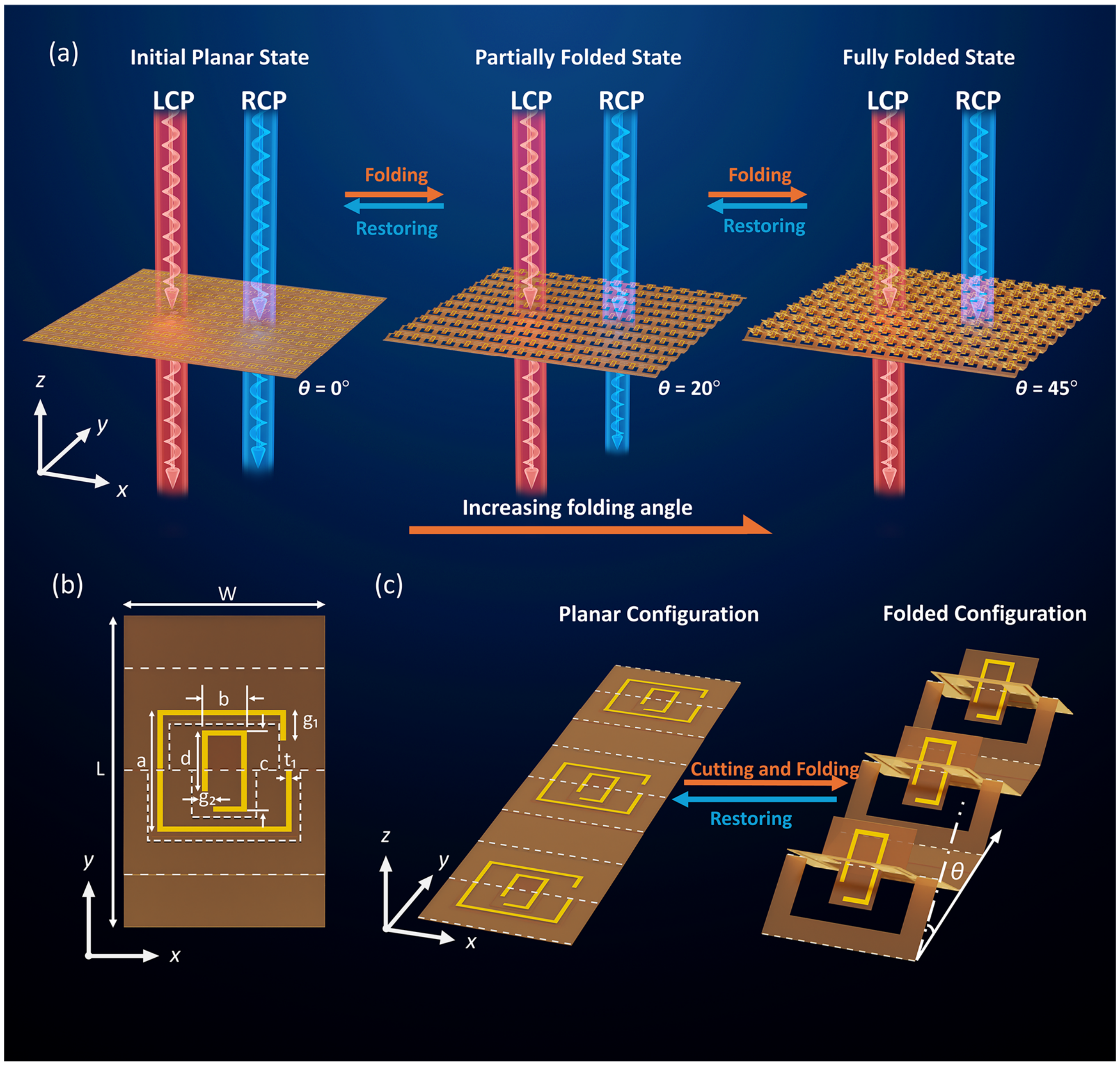

**Fig. 1 | Schematic illustration of the kirigami-based reconfigurable metasurface.** (a) Initially, in the planar state, the metasurface is achiral. With the kirigami method, that is, cutting and folding of the metasurface, it becomes a chiral metasurface with asymmetric transmissions for the left- (LCP) and right-circularly polarized (RCP) incidences. When the folding angle $\theta$ = 45°, a near-unity circular dichroism (CD) is achieved. (b) Top view of a unit cell of the initial planar state of the metasurface. (c) The reconfigurable mechanical deformation of the kirigami-based metasurface.

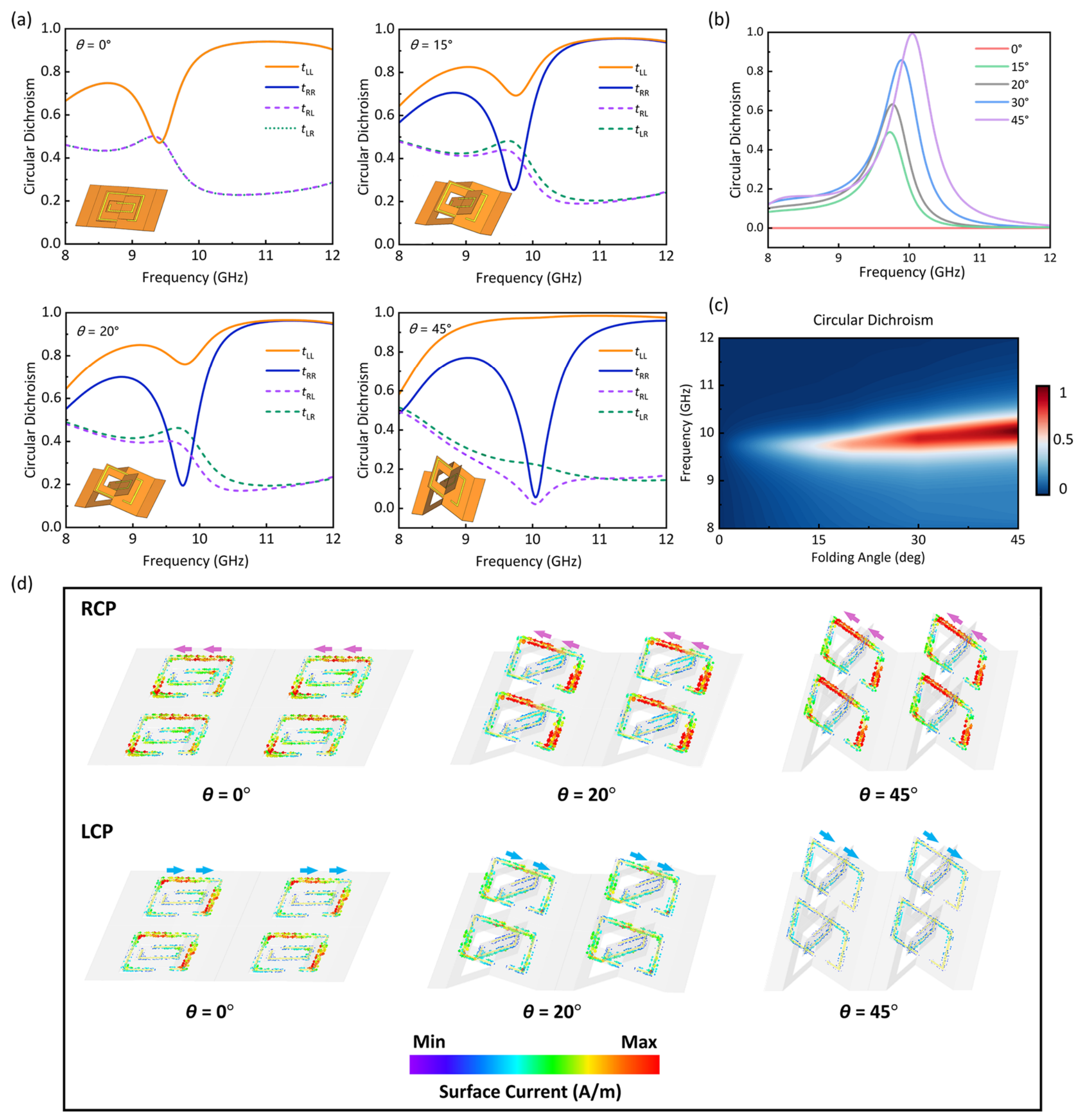

**Fig. 2 | Transmission and chiral responses of the reconfigurable metasurface under different folding angles.** (a) The simulated spectra of transmission coefficients of the metasurface under normal incidences of LCP and RCP waves at specific folding angles of 0°, 15°, 20°, and 45°, respectively. (b) The calculated CD spectra at the corresponding folding angles, showing that CD generally increases with the deformation. (c) The CD map as a function of frequency and folding angle, revealing a strong correlation between folding-induced asymmetry and enhanced chiral response centered around 10 GHz. (d) The simulated surface current distributions at

folding angles of 0°, 20°, and 45°, under LCP and RCP excitation at resonance frequency.

To verify this mechanism, we use the commercial finite element software COMSOL Multiphysics to simulate the circular polarization transmissions of the metasurfaces with different folding angles, as shown in Fig. 2(a). Here, $t_{ij}$ (*i, j* = *L*, *R*; *L* & *R* standing for LCP and RCP, respectively) denotes the transmission coefficient under circular polarization from the state *i* to state *j*. For the unfolded metasurface, the simulated co-polarized transmission coefficients spectra $t_{RR}$ and $t_{LL}$ are nearly identical, indicating near-zero intrinsic chirality. As the folding angle $\theta$ increases (e.g., from 10° to 20°), a growing difference emerges between $t_{LL}$ and $t_{RR}$. At $\theta$=45°, $t_{LL}$ approaches near 1 while $t_{RR}$ drops close to zero, demonstrating a high-efficiency circular polarization-selective transmission and implying a significant CD. Here, we quantify the CD as[43]:

$$CD = \frac{t_{LCP} - t_{RCP}}{t_{LCP} + t_{RCP}} = \frac{\left(\left|t_{LL}\right|^2 + \left|t_{RL}\right|^2\right) - \left(\left|t_{RR}\right|^2 + \left|t_{LR}\right|^2\right)}{\left(\left|t_{LL}\right|^2 + \left|t_{RL}\right|^2\right) + \left(\left|t_{RR}\right|^2 + \left|t_{LR}\right|^2\right)}, \tag{1}$$

where $t_{LCP(RCP)} = |t_{LL(RR)}|^2 + |t_{RL(LR)}|^2$ represents the total transmission for LCP (RCP) incidence. The simulated CD spectra of the metasurface at folding angles of 0°, 15°, 20°, 30°, and 45° are plotted in Fig. 2(b), showing that the chiral response intensifies with increasing folding angle. At $\theta$ = 45°, a significant CD peak of 0.90 is observed near the resonance frequency of 10.1 GHz. This nearly monotonic trend is further illustrated in Fig. 2(c), which plots the CD value versus a continuous folding angle $\theta$.

Although the kirigami metasurface undergoes a mechanical deformation in the relative position of the SRRs, which slightly alters the effective current paths and coupling configurations, the resonant eigenmode relevant to the chirality remains highly localized due to its nearly flat band (see Supporting Information Note S2). As a result, the resonance frequency in the CD response experiences only a slight shift even for a large mechanical deformation ($\theta$ = 45°), demonstrating the robustness of the intrinsic chirality.

To further investigate the mechanism behind the chirality enhancement induced by structural deformation, Fig. 2(d) presents the simulated surface current distributions under LCP and RCP excitation at different folding angles. At $\theta$ = 0°, the structure remains planar, and the current distributions induced by LCP and RCP incidence are nearly symmetric, resulting in a negligible chiral response. As the folding angle increases to 20° and 45°, the degree of freedom in $z$ direction emerges, and the current distributions become increasingly asymmetric between the LCP and RCP incidences, forming distinct current pathways. This behavior corresponds well with the growing disparity between the $t_{RR}$ and $t_{LL}$ components observed in the transmission spectra (Fig. 2(a)). Meanwhile, under LCP and RCP incidences, the multipole responses dominated by the surface current of the SRRs are different (especially the asymmetric response strength of the toroidal and electrical dipole moments), which is the main source of the strong intrinsic chirality, as will be further discussed later.

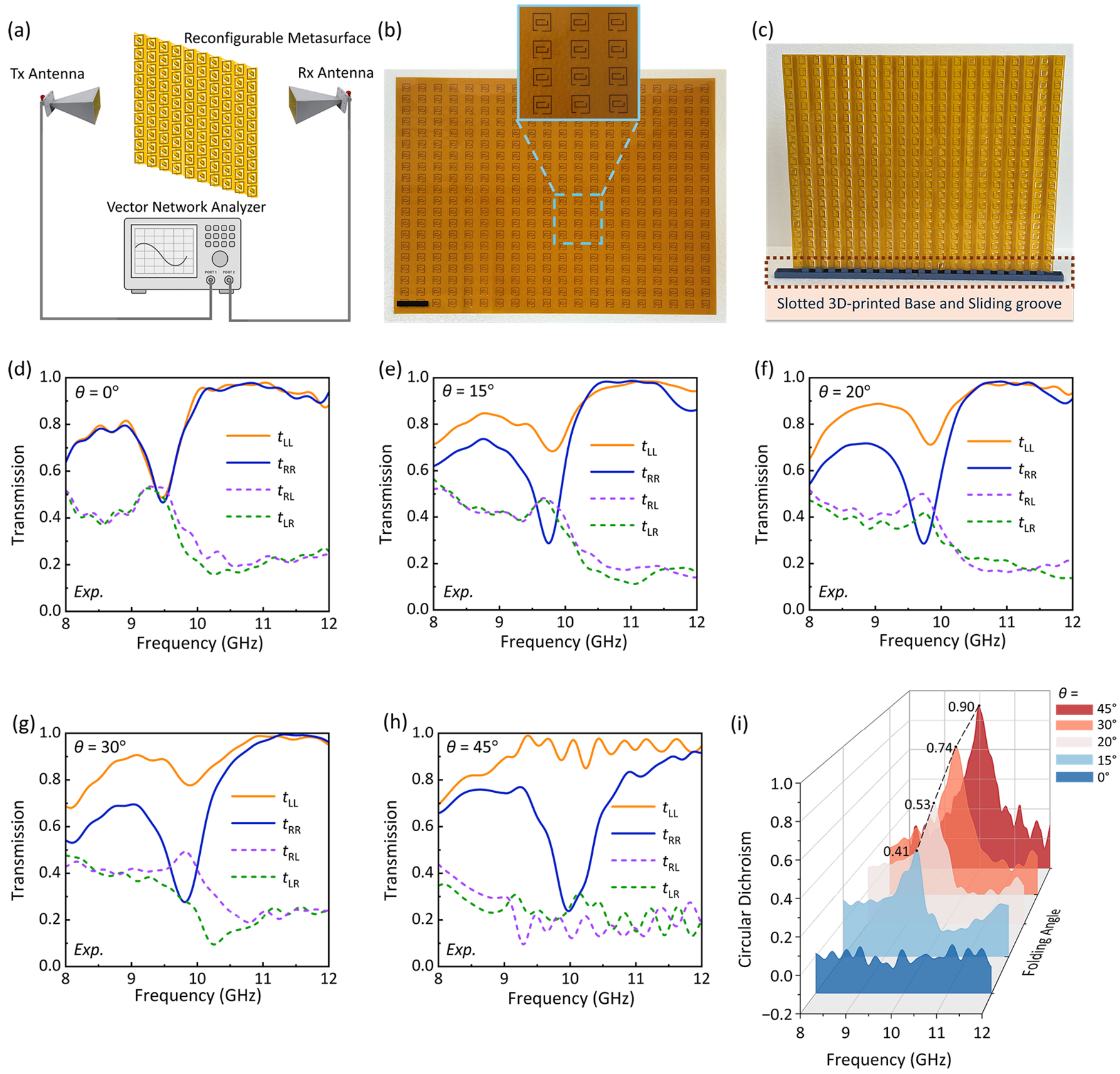

**Fig. 3 | Experimental demonstration of the reconfigurable chiral metasurface.** (a) Schematic of the experimental setup using two linear-polarized horn antennas for transmission measurement, along with a vector network analyzer. (b)-(c) Photographs of the fabricated metasurface sample (b), with an inset showing zoom-in details of the unit cells, and its folding configuration (c). The sample is supported from the bottom by a slotted 3D-printed base integrated with sliding grooves, which enable adjustable folding. The black scale bar in (b) is 50 mm. (d)-(h) The experimentally measured transmission coefficient spectra of circular polarization components at folding angles of 0°, 15°, 25°, 30°, and 45°, respectively. (i) The measured CD spectra calculated

from (d)-(h) with peak value reaching 0.90 near 10 GHz.

To experimentally validate the simulation results, we fabricated the metasurface based on our design using flexible printed circuit (FPC) technology, with the desired patterns precisely cut by femtosecond laser processing. The photographs of the fabricated metasurface sample, including both an uncut/unfolded sample and an enlarged view, are shown in Fig. 3(b). The metasurface consists of an array of 20 × 20 unit cells. The assembled structure and mounting of the reconfigurable chiral metasurface are also illustrated. During the folding process, intermediate planar regions naturally form between the periodically arranged unit cells. Hardened foam strips are attached to the backside of these planar connection regions to provide flexible support and constrain deformation, as shown in Fig. 3(c) (see Supporting Information Note S3 for details of the tunable support platform), which enables a continuous control of the folding angle of the metasurface.

The experimental setup used to measure the transmission coefficients is presented in Fig. 3(a). The system consists of a pair of linear polarized horn antennas, working as transmitter (Tx) and receiver (Rx), which are connected to a vector network analyzer (VNA) for extraction of the $S_{21}$ parameters as the transmission coefficients (see Supporting Information Note S4). The metasurface sample is placed between the two antennas to measure its chiral responses of different folding configurations. To obtain the transmission coefficients for circular polarization, we first measured the four linear transmission components—$t_{xx}$, $t_{yy}$, $t_{xy}$, and $t_{yx}$. We then converted them into circular polarization components using the Jones matrix formalism[44-45] (see Supporting Information Note S5), as expressed below:

$$\begin{pmatrix} t_{RR} & t_{RL} \\ t_{LR} & t_{LL} \end{pmatrix} = \frac{1}{2}\begin{pmatrix} t_{xx}+t_{yy}+i\left(t_{xy}-t_{yx}\right) & t_{xx}-t_{yy}-i\left(t_{xy}+t_{yx}\right) \\ t_{xx}-t_{yy}+i\left(t_{xy}+t_{yx}\right) & t_{xx}+t_{yy}-i\left(t_{xy}-t_{yx}\right) \end{pmatrix}. \tag{2}$$

As shown in Fig. 3 (d)-(h), we experimentally obtained the circular polarized transmission spectra at folding angles of 0°, 15°, 20°, 30°, and 45°, respectively. The measured results are in good agreement with full-wave simulation data. Fig. 3(i) shows the experimental CD spectra as functions of frequency and folding angle, revealing a gradual increase in CD values with the increase of folding angle and reaching a peak value of 0.90 in experiments. The results soundly confirm the strong tunability of the chiral response of our kirigami-based metasurface enabled by simple cutting and folding.

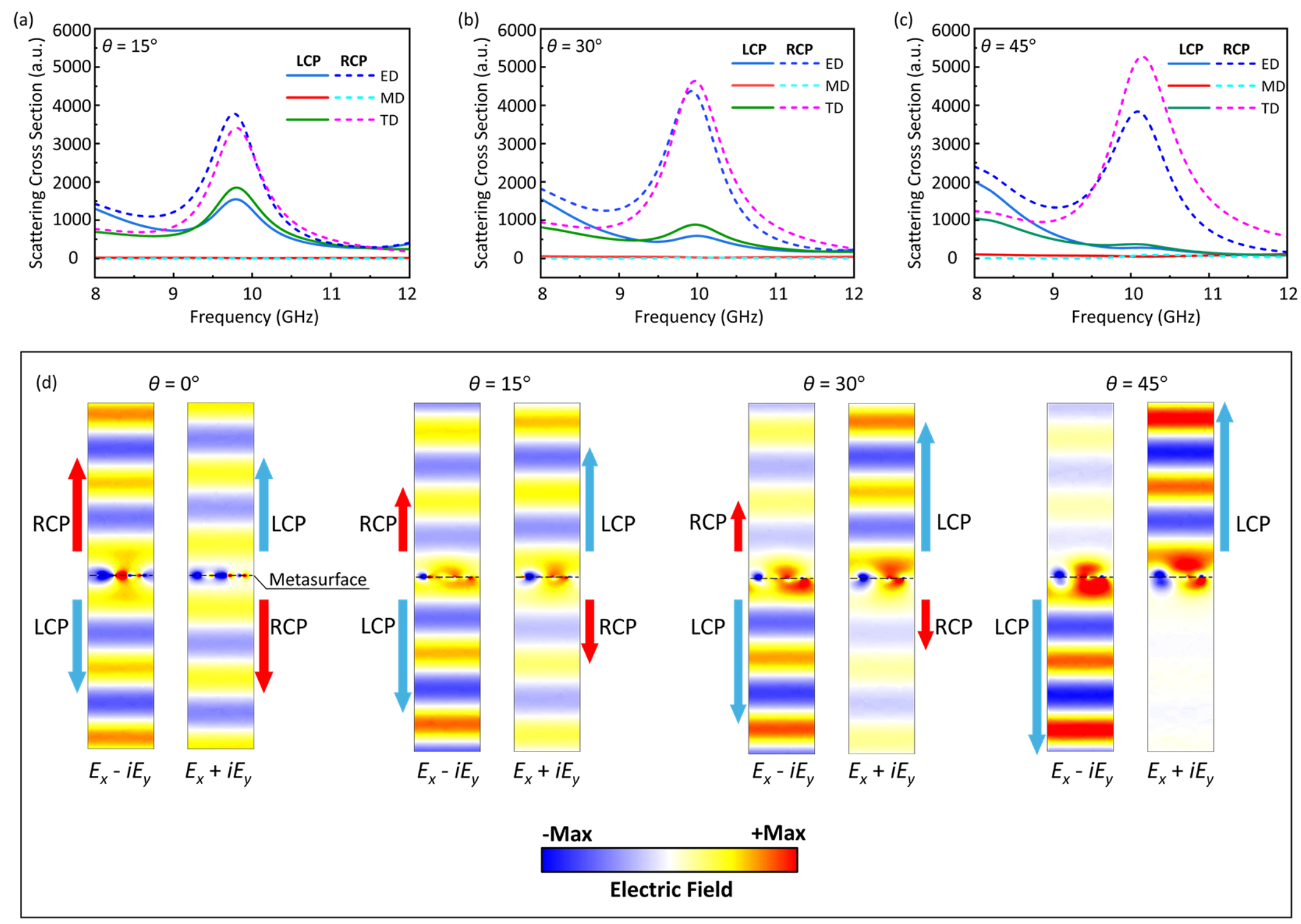

**Fig. 4 | Multipolar analysis of chiral response and asymmetric field distribution in eigenmodes.** (a)-(c) The calculated scattering cross sections of ED, MD, and TD components under LCP and RCP incidences at different folding angles of 15°, 30°, and 45°. Compared to the LCP case, the RCP incidence induces significantly stronger ED and TD contributions near resonance, indicating a polarization-sensitive multipole response. (d) The simulated electric field distributions of the eigenmode at folding angles of 0°, 15°, 30°, and 45°. The circular polarized components $E_x \pm iE_y$ are plotted. As the folding angle $\theta$ increases, the mirror symmetry with respect to the *xy*-plane is progressively broken, and the asymmetric circular-polarized radiation of the eigenmode becomes increasingly pronounced.

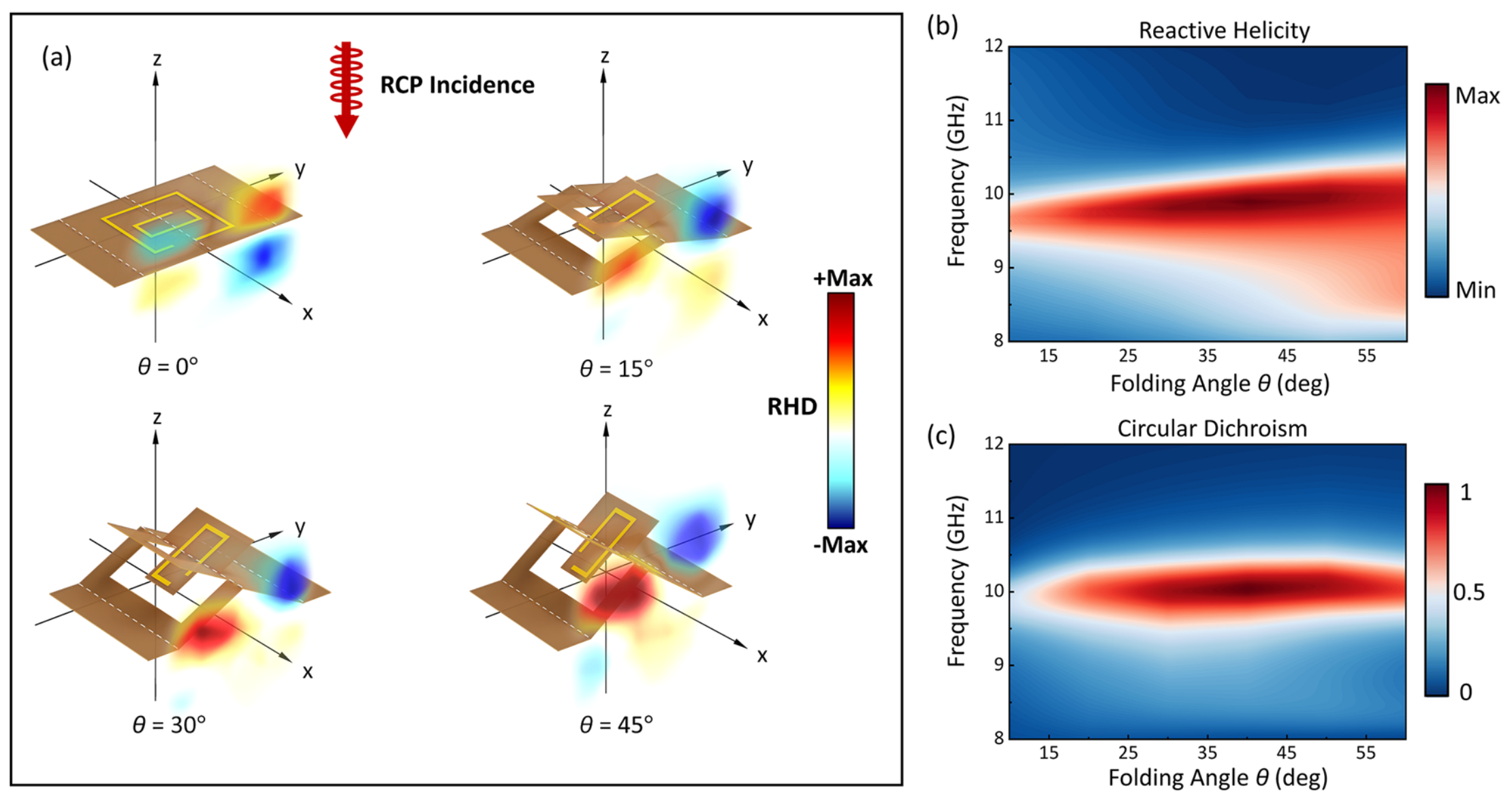

**Fig. 5 | Angle-dependent evolution of intrinsic chiral near-fields and comparison with CD responses.** (a) The simulated RHD distributions under RCP incidence at folding angles of 0°, 15°, 30°, and 45°. The RHD fields exhibit progressively stronger asymmetry in the *yz*-plane as the folding angle $\theta$ increases, indicating the emergence and enhancement of intrinsic chiral near-fields (blank regions: zero values where **B** ⊥ D). The calculated total reactive helicity (b) and CD (c) as functions of folding angle and frequency, showing a strong correlation between structural deformation and chiral response.

To further elucidate the physical mechanism underlying the intrinsic chirality of the metasurface, we calculated the scattering cross sections of the multipoles (see Supporting Information Note S6) excited by a circular polarized wave of the reconfigurable chiral metasurface at folding angles of 15°, 30°, and 45°. The analysis includes the electric dipole (ED) moment **P**, magnetic dipole (MD) moment **M**, and toroidal dipole (TD) moment **T**. The scattering cross sections contributed by each dipole moment are defined as follows[46-49]:

$$C_{sca}^{(ED)} = \frac{2\omega^4}{3c^3}\left|\mathbf{P}\right|^2, \tag{3}$$

$$C_{sca}^{(MD)} = \frac{2\omega^4}{3c^3}\left|\mathbf{M}\right|^2, \tag{4}$$

$$C_{sca}^{(TD)} = \frac{4\omega^5}{3c^4}\mathrm{Im}\left(\mathbf{P}^* \cdot \mathbf{T}\right) + \frac{2\omega^6}{3c^5}\left|\mathbf{T}\right|^2. \tag{5}$$

Fig. 4(a)-(c) show the calculated results under LCP and RCP incidences. Under LCP incidence, both the TD and ED moments exhibit relatively weak multipole responses, which further decrease as the folding angle increases. In contrast, under RCP incidence, the excitations of TD and ED moments become significantly stronger with increasing folding angle, especially in the 9.7–10.1 GHz frequency range. In both cases, however, the MD moment response remains negligible for all angles across the frequency range. All the components of TD and ED ($T_x$, $T_y$, $T_z$ and $p_x$, $p_y$, $p_z$) are calculated, with detailed results shown in Supporting Information Note S7＆S8. A further evaluation of the far-field scattering characteristics based on the total electric dipole moment $\mathbf{p}-ik\mathbf{T}$, which captures the combined contribution of ED and TD moments, is provided in Note S9 of the Supporting Information. Generally, the multipole expansion results mirror the surface current distributions. In essence, this asymmetric response is owing to the non-coplanar SRRs, which give rise to a chiral eigenmode, primarily dominated by ED and TD moments with both in-plane and out-of-plane components. On the other hand, in the no-folding case ($\theta$ = 0°), the coplanar SRRs of the metasurface give rise to an achiral eigenmode dominated only by in-plane ED and TD moments (see Supporting Information Note S10).

This behavior of asymmetric responses is further elucidated by the far-field radiation from the eigenmode shown in Fig. 4(d), which depicts the circular polarized electric field components of $E_x \pm iE_y$ at various folding angles of the metasurface. At $\theta$ = 0°, the circular polarized radiation on both sides remains symmetric, indicating the LCP and RCP incidences will be coupled symmetrically to the metasurface, also meaning that their transmissions or reflections will be essentially achiral. As the *xy*-plane mirror symmetry of the metasurface is progressively broken with the increasing folding angle $\theta$, the circular polarized radiation from the eigenmode exhibits increasingly asymmetric distributions on both sides of the metasurface, and a nearly unidirectional radiation pattern is achieved at $\theta$ = 45°. In fact, the increasing out-of-plane ED and TD components with respect to the increasing folding angle $\theta$ give rise to the asymmetric radiation of the LCP and RCP waves from the metasurface. Reciprocally, the asymmetric radiation corresponds to an asymmetric coupling between the metasurface and the LCP and RCP incidences, which in turn leads to a significant intrinsic chirality around the frequency of the eigenmode.

To further quantify the near-field origin of the chirality of the metasurface at different folding angles, we introduce the RHD[50] as an evaluation metric, which plays a significant role in characterizing the generation of chiral radiation. The RHD is

defined as:

$$\chi_r = \frac{c}{\omega}\mathrm{Re}\left(\mathbf{B}^* \cdot \mathbf{D}\right), \tag{6}$$

where $\omega$ is the angular frequency of the EM wave, **B** and **D** are the magnetic flux density and electric displacement, respectively, and $^*$ denotes the complex conjugate.

As shown in Fig. 5(a), the simulated RHD distributions under RCP incidence reveal the evolution of near-field chirality with folding angle $\theta$. This behavior maps onto the multipolar analysis presented earlier. At $\theta$ = 0°, the RHD field is symmetrically distributed and balanced about the *xy*-plane, implying a net-zero helicity and achiral response in the far field. As the folding angle $\theta$ increases, a significantly asymmetric RHD distribution gradually emerges, which is accompanied by a gradually increasing out-of-plane TD component, leading to significant helicity and chiral responses. The RHD field of the eigenmode also shows a similar evolution trend, transitioning from a symmetric distribution at $\theta$ = 0° to increasingly asymmetric as the folding angle $\theta$ increases (see Supporting Information Note S11).

By integrating the RHD distribution around the metasurface structure under the RCP incidence, we obtain the total reactive helicity for different configurations to quantify the asymmetry and chirality. As shown in Fig. 5(b), the reactive helicity increases markedly with the folding angle $\theta$, and its trend closely matches that of the CD value shown in Fig. 5(c). Our results thus confirm that the folding angle $\theta$ effectively controls the asymmetry in the RHD distribution, and the resulting chiral response of the kirigami-based metasurface, as quantified by the reactive helicity and CD values. In fact, the asymmetric RHD distribution leads to asymmetric coupling for the folded metasurface with LCP and RCP incidences, giving rise to the intrinsic chirality. Therefore, the intrinsic chirality can be reversed by flipping the folding direction of the kirigami-based metasurface, which in turn reverses its asymmetric coupling with LCP and RCP incidences (see Supporting Information Note S12).

Moreover, we note that the folding can also lead to asymmetric chirality for the kirigami-based metasurface under oblique incidences, as schematically depicted in the upper panel in Fig. 6(a). As is known, when there is no folding, the CD spectra due to extrinsic chirality are anti-symmetric with respect to the oblique angles of incidence, as shown in Fig. 6(b) and experimentally verified in Fig. 6(d). However, for small angles of incidence such as $\alpha = \pm 10°$, as shown in Fig. 6(c), numerically calculated CD spectra reveal that its total chirality has become extremely asymmetric. As illustrated in the lower panel in Fig. 6(a), if the extrinsic chirality induced by angle of incidence and the intrinsic chirality folding angle “cancel” each other, the total chirality will become very weak, while for the other case, the total chirality will be significantly enhanced. The corresponding results are also experimentally validated, as shown in Fig. 6(e), where the measured CD spectra directly reveal the extremely asymmetric chirality. In other words, by tuning the folding angle of the metasurface, we can also realize active control of circular dichroism at oblique incident angles through intrinsic chirality (see Supporting Information Note S13). Therefore, we establish a direct and tunable connection between the two types of chirality within a single platform by leveraging the kirigami-based patterns.

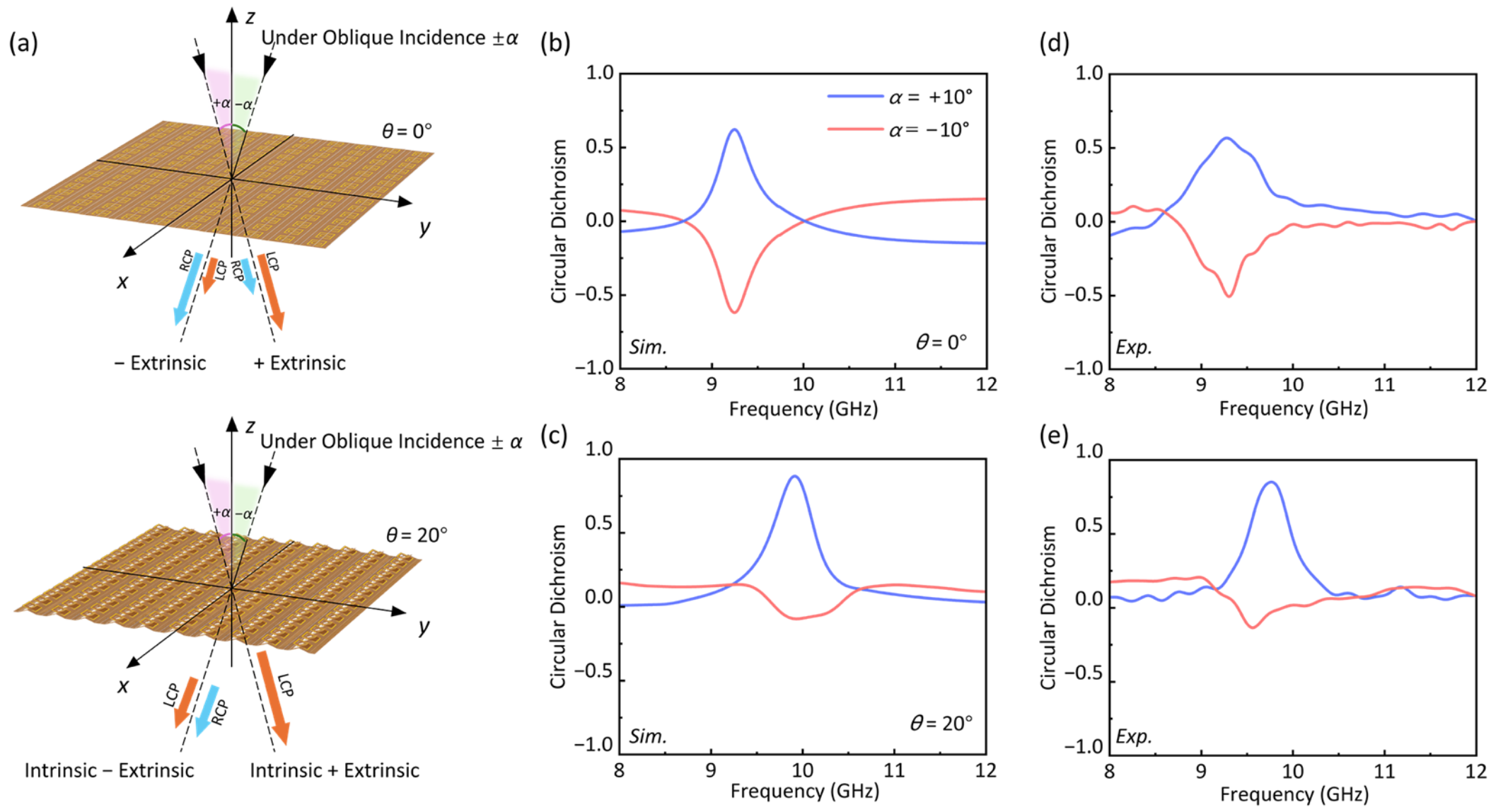

**Fig. 6 | Angle-dependent asymmetric chirality.** (a) Schematic of the chirality of the kirigami-based metasurface, anti-symmetric when there is no folding (only extrinsic chirality), and asymmetric where there is a folding (intrinsic chirality alongside extrinsic chirality). (b, c) Simulated and (d, e) experimentally measured CD spectra representing chirality under opposite angles of incidence $\pm 10°$, when there is no folding (b, d) and a folding angle ($\theta$ = 20°). The folding can lead to extremely asymmetric chirality (c, e).

## 3. Conclusion

In conclusion, we have proposed and systematically demonstrated a reconfigurable metasurface based on kirigami for continuous modulation of intrinsic chirality. By applying a simple cutting and folding technique, an initially planar, achiral metasurface is transformed to exhibit tunable intrinsic chirality. Experimental validation through microwave measurements confirms its exceptional CD under normal incidence, with the CD spectra varying smoothly with the folding angle and the peak CD values gradually reaching up to 0.9. Multipole analysis further clarifies that the strong chiral response primarily arises from non-planar electric and toroidal dipole moments induced by folding. The metasurface, in its planar and achiral state, also supports strong extrinsic chirality under oblique incidence.

Therefore, our kirigami approach integrates extrinsic and intrinsic chirality within a single metasurface platform, delivering robust and tunable chiral responses. Its thin, lightweight structure and dynamic chirality control through mechanical reconfiguration highlight the kirigami metasurface's practical feasibility and versatility. This approach is also largely independent of specific material properties, enabling applications across higher frequency regimes, such as terahertz and infrared. With its remarkable efficiency, structural adaptability, and exceptional performance, this design strategy offers transformative potential for advancing cutting-edge research and pioneering applications in photonic and optoelectronic technologies, particularly in chiral sensing, polarization encoding, holographic display, and a broad spectrum of related interdisciplinary fields.

## Methods

### Sample Fabrication

The samples were fabricated using flexible printed circuit (FPC) technology. A 0.1 mm thick polyimide (PI) film with a dielectric constant of $\varepsilon_r = 3.5$ was used as the substrate. A 35 μm thick copper layer was deposited onto one side of the PI film through electroplating, forming a uniform conductive surface for subsequent patterning. The metasurface pattern was defined using standard photolithography followed by wet etching. To precisely form the dashed fold slits and kirigami-patterned cuts between unit cells, femtosecond laser micromachining was employed, enabling high-accuracy cutting without thermal damage.

**Sample Reconfiguration**

To enable mechanical reconfiguration, the fabricated metasurface was mounted on a custom-designed tunable support platform. The metasurface sample contains flat connection regions between adjacent unit cells, where hardened foam strips were bonded to provide flexible mechanical support and restrict undesired deformation. The bottom side of each foam strip was embedded into a slotted 3D-printed base, ensuring stable alignment. This base was further inserted into a long sliding groove, allowing the entire base–foam strip–sample assembly to move freely along the groove. This nested structure enables smooth and continuous adjustment of the folding angle of the metasurface. See Supporting Information Note S3 for details.

**Microwave Measurement**

All measurements were conducted in a microwave anechoic chamber. Two sets of linearly polarized horn antennas (operating within the frequency range of 1–18 GHz) were connected to Port 1 and Port 2 of a vector network analyzer (VNA 3674H，Ceyear) via coaxial cables, serving as the transmitter and receiver, respectively. To satisfy the far-field condition, the distance between the antennas and the metasurface was kept sufficiently large; however, an excessively long separation would result in pronounced diffraction effects due to the limited size of the metasurface, thereby compromising measurement accuracy.

Before the measurement, the two horn antennas were aligned at the same height and directly facing each other. With the sample being absent and only foam supports and mounting structures, full calibration of the VNA and connecting cables was performed. Co-polarized complex transmission coefficients ($t_{xx}$, $t_{yy}$) were measured and obtained from $S_{21}$ parameters with parallel alignment of transmitting and receiving antennas. Cross-polarized complex transmission coefficients ($t_{xy}$, $t_{yx}$) were obtained under orthogonal polarization alignment.

**Simulation Settings**

All full-wave numerical simulations were performed using the finite element method (FEM) implemented in the commercial software COMSOL Multiphysics. The metasurface unit cell was embedded in a homogeneous background medium (air). Two interleaved SRRs were modeled as perfect electric conductor (PEC) surfaces, while the substrate was defined as a 0.1 mm thick polyimide layer with a relative permittivity of 3.5 and a loss tangent of 0.01. Non-uniform meshes were applied, with the largest element size kept below 1/5 of the minimum operating wavelength to ensure numerical accuracy. To eliminate spurious reflections from the simulation domain boundaries, two perfectly matched layers (PMLs) were introduced at the top and bottom surfaces. The incident electromagnetic wave propagated along the $z$-axis. Periodic boundary conditions were applied along the $x$ and $y$ directions, while port boundary conditions were assigned to the top and bottom surfaces to excite the wave and extract the $S_{21}$ parameters.

**Supporting Information**

See Supporting Information for supporting content.

**Data Availability**

The data that supports the findings of this study are available from the corresponding authors upon reasonable request.

**Author Contributions**

Y. Yao and X. Wu conceived the idea and initiated the project. Y. Yao and A. Luo constructed the simulation model and performed numerical calculations. Y. Yao, H. Li, and S. Kang set up the experiment, conducted the measurements, and performed initial data processing. All authors participated in the discussion of the results, and Y. Yao led the writing of the manuscript, with input and revisions from all co-authors. X.

Xia, H. Li, and X. Wu supervised the project and contributed to all aspects of analysis and experimental design.

## Acknowledgements

This research was supported by the National Natural Science Foundation of China (No. 12304348, 62505236), Guangdong Basic and Applied Basic Research Foundation (No. 2025A1515011470), Guangdong University Featured Innovation Program Project (2024KTSCX036), Guangdong Provincial Project (2023QN10X059), Guangzhou-HKUST(GZ) Joint Funding Program (2025A03J3783), Guangzhou Municipal Science and Technology Project (2024A04J4351), Guangzhou Young Doctoral Startup Funding (2024312028), and the Fundamental Research Funds for the Central Universities (WUT:104972024RSCbs0023). The authors would like to acknowledge Wave Functional Metamaterial Research Facility (WFMRF) of The Hong Kong University of Science and Technology (Guangzhou) for the experimental support.